\documentclass[11pt]{article}
\usepackage{graphicx}

\def\singleandabitspaced{\baselineskip=\normalbaselineskip\multiply
    \baselineskip by 110\divide\baselineskip by 100}
\def\singlespaced{\baselineskip=\normalbaselineskip}

\newcommand{\centeron}[2]{{\setbox0=\hbox{#1}\setbox1=\hbox{#2}\ifdim
                             \wd1>\wd0\kern.5\wd1\kern-.5\wd0\fi \copy0
                             \kern-.5\wd0\kern-.5\wd1\copy1\ifdim\wd0>\wd1
                             \kern.5\wd0\kern-.5\wd1\fi}}
\newcommand{\ltap}{\>\centeron{\raise.35ex\hbox{$<$}}
                     {\lower.65ex\hbox{$\sim$}}\>}
\newcommand{\gtap}{\>\centeron{\raise.35ex\hbox{$>$}}
                     {\lower.65ex\hbox{$\sim$}}\>}
\newcommand{\gsim}{\mathrel{\gtap}}
\newcommand{\lsim}{\mathrel{\ltap}}

\newcommand{\ms}{M_S^{}}

\newcommand{\md}{M_D^{}}

\newcommand{\xmin}{x_{min}}

\def\gev{\rm GeV} 
\def\tev{\rm TeV} 

\begin{document}

\singlespaced

\begin{titlepage}

\vspace*{-16mm}
\begin{flushright}
{\small
MADPH-04-1392 \\
hep-ph/0408348  \\
}
\end{flushright}

\begin{center}
\vspace*{0.8in}
\mbox{\Large \textbf{The Particle Physics Reach of High-Energy Neutrino Astronomy}} \\
\vspace*{1.6cm}
{\large Tao Han$^1$ and Dan Hooper$^2$} \\
\vspace*{0.5cm}
{\it $^1$ Physics Dept., Univ.  of Wisconsin, 
1150 University Ave., Madison, WI 53706, USA\\ 
$^2$ Astrophysics Dept., Univ. of Oxford, 
Denys Wilkinson Laboratory, OX1 3RH Oxford, UK} \\
\vspace*{0.6cm}
{\tt than@physics.wisc.edu, hooper@astro.ox.ac.uk} \\
\vspace*{1.5cm}
\end{center}

\begin{abstract} 
\singleandabitspaced

We discuss the prospects for high-energy neutrino astronomy to study 
particle physics in the energy regime comparable to and beyond that
obtainable at the current and planned colliders.  We describe the various 
signatures of  high-energy cosmic neutrinos expected in both neutrino telescopes 
and air shower experiments and discuss these measurements 
within the context of theoretical models with a quantum gravity or
string scale near a TeV, supersymmetry and scenarios with interactions induced 
by electroweak instantons. We attempt to access the particle 
physics reach of these experiments.

\end{abstract}

\end{titlepage}

\newpage
\setcounter{page}{2}

\singleandabitspaced

\section{Introduction}

In recent endeavors, explorations of high-energy 
particle physics have largely been conducted in accelerator
 experiments, and for good reason. Accelerator laboratories provide controlled, 
high-luminosity
 environments in which very precise levels of measurement can be reached. Despite these
 advantages, astrophysics experiments have also revealed a great deal of particle physics 
beginning with Anderson's discovery of the positron in 1932, then the muon, the pion etc.,
predating accelerator experiments,  
and continuing to the observation of neutrino masses and mixings. 
It is clear that astrophysics 
has much to offer in studying the fundamental aspects of particle physics.

Particle  physics has entered an exciting era. The Standard Model (SM) of 
the strong and electroweak interactions has been experimentally verified to high 
precision while the mechanisms for electroweak symmetry breaking
and mass generation remain largely unknown. Theoretical arguments and indirect 
experimental evidence imply the existence of new physics near the electroweak scale
below a few TeV  \cite{Altarelli:2003rm}. 
The leading candidates of theoretical models beyond the Standard Model include weak scale
supersymmetry (SUSY) \cite{Martin:1997ns}, 
strongly interacting dynamics \cite{tcs}, and low-scale
string or quantum gravity \cite{add,rs}. It is encouraging that 
all of the above scenarios often
lead to observable signatures in next generation colliders such as the CERN
Large Hadron Collider (LHC) and an $e^+e^-$ linear collider.
 
The field of high-energy neutrino astronomy finds itself in a position to contribute to two very different areas of science: astronomy and particle physics  \cite{review}.
On the one hand, the next generation of
neutrino telescopes may reveal the origins of the highest-energy cosmic rays, help us understand the progenitors of gamma-ray bursts, and provide other insights into some of the greatest outstanding astrophysical puzzles. On the other hand, very high-energy cosmic neutrinos present a unique opportunity to study the interactions of elementary particles at energies comparable to and
beyond  those obtainable in current or planned colliders.  
This is the main advantage of such experiments over traditional collider experiments.  Currently, the highest energy achieved in
collider experiments is at FERMILAB's Tevatron, with $E_{CM} \approx 2$ TeV. 
This center-of-mass energy roughly corresponds to a PeV neutrino striking a nucleon 
at rest, $E_\nu=E_{CM}^2/2m_N^{}$. 
Even the LHC at CERN will only reach energies which correspond to 100 PeV 
cosmic neutrinos. It is certainly plausible, as we will discuss, 
that there is a neutrino flux at energies well beyond 1 EeV.
Even crude measurements of neutrino cross sections at extremely high energies would 
provide powerful tests of fundamental physics at and beyond a scale of $1-10$ TeV. Additionally, 
sources of high-energy neutrinos may be observed from distances of hundreds or thousands of megaparsecs, providing baselines for tests of neutrino oscillations or decays which could not be carried out using accelerator, atmospheric or solar neutrinos. 

The experimental status of high-energy neutrino astronomy is  developing rapidly. Current technologies such as the AMANDA-II \cite{amanda} and RICE \cite{rice} experiments at the South Pole have proven successful, but with too little sensitivity to reach many of the most interesting physics goals. Several new experiments are soon to enter the field fortunately. 
IceCube \cite{icecube} will expand the effective area of AMANDA-II by  more than a factor
of 20 while also improving both the angular and energy resolution. ANTARES \cite{antares}, in the Mediterranean, will use a similar technique, but with sensitivity to neutrino induced muons of lower energy (down to 10 GeV). Radio techniques will be employed in the balloon-based ANITA \cite{anita} experiment which has it's first flight scheduled in the next year or two. High-energy cosmic ray experiments, such as the Pierre Auger observatory \cite{auger} and space-based observatories such as OWL \cite{owl} or EUSO \cite{euso}, will also be sensitive to ultrahigh-energy cosmic neutrinos. Other proposals, such as using acoustic techniques \cite{acoustic} or natural salt domes as a Cerenkov medium \cite{salsa}, or expanding IceCube into a multi-kilometer ultrahigh-energy experiment \cite{hypercube}, have been discussed as well. With the development of these many and varied techniques, a new window into fundamental particle physics 
will be opened \cite{review}, possibly predating the LHC experiments.

This article is a mini review on the potential of high-energy neutrino astronomy in studying particle physics beyond the Standard Model. 
We first present the current theoretical predictions of the high-energy
cosmic neutrino flux in Sec.~\ref{flux}. We then discuss methods for exploring particle
physics via high-energy neutrino astronomy in Sec.~\ref{part}, 
paying particular attention to future neutrino telescopes 
and air shower observatories.  We summarize in Sec.~\ref{theory}
the predicted experimental signatures from various new physics scenarios,
including models with low-scale quantum gravity, low-scale string resonances, 
black holes and $p$-branes, electroweak instantons, and supersymmetry. 
We draw our conclusions in Sec.~\ref{conclude}.

\section{The High-Energy Cosmic Neutrino Flux}
\label{flux}

Just as the performance of an accelerator experiment crucially 
depends on its luminosity, the particle physics reach of high-energy neutrino astronomy will depend on the incoming flux of cosmic neutrinos.  Here we will briefly review some of the arguments for various neutrino fluxes from cosmic accelerators.

\begin{figure}[t]
\centering\leavevmode
\includegraphics[width=3.5in]{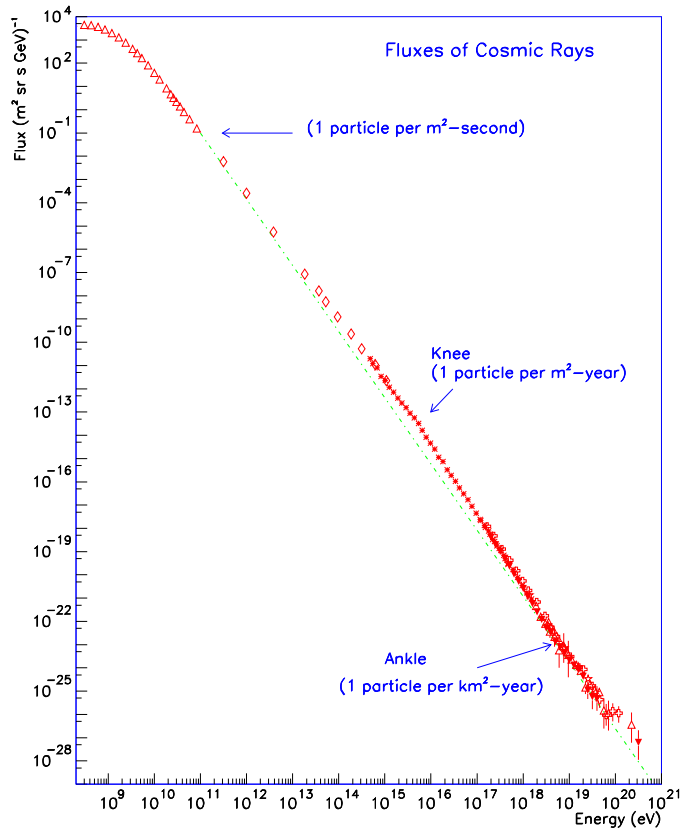}
\caption{The observed cosmic ray spectrum.}
\label{rays}
\end{figure}

The spectrum of cosmic rays has been well measured up to energies near 
$10^{20}$ eV  ($10^{8}$ TeV in the lab frame) 
where the experiments become limited by poor statistics. The spectrum consists of a series of power laws which change at energies called to as the ``knee'' and ``ankle'' (see Fig.~\ref{rays}). The standard process believed to be responsible for the observed spectrum of cosmic rays is the acceleration of charged particles via second-order Fermi acceleration \cite{fermis}. 
In Fermi's original paper on the subject, he proposed that cosmic rays were accelerated by 
reflecting off of time-varying magnetic fields associated with galactic clouds moving with randomly distributed velocities. Although this process will statically accelerate charged particles, it does so very slowly, and is generally not capable of even countering the effects of energy losses by ionization and other processes. If instead we consider a compact region of dense plasma, however, the random motion of the matter and associated magnetic fields can be sufficient to accelerate charged cosmic rays to very high energies. Although there is no strong evidence as of yet, it is likely that supernova remnants accelerate most of the cosmic rays up to the knee in the spectrum, occurring around $10^{15}$ eV, by this mechanism. A generic feature of Fermi acceleration is a power-law spectrum, $dN/dE \propto E^{-\alpha}$, where $\alpha \simeq 2$ \cite{alpha2}.

The maximum energy to which a cosmic ray source may accelerate particles can be estimated by a simple argument. First, we assume that to accelerate a proton to a given energy in a given magnetic field, the size of the accelerator must be larger than the gyroradius of the particle's orbit:
\begin{equation}
R > R_{\rm{gyro}} = \frac{E}{B}.
\end{equation}
This condition yields a maximum energy of
\begin{equation}
E_{\rm{max}} = \gamma B R,
\end{equation}
where $\gamma$ is the Lorentz factor of the cosmic accelerator. To produce cosmic rays with energies near the highest observed ($\sim 10^{20}$ eV), very compact objects are required. For the most compact objects, we can consider $R \sim G M/c^2$, or the Schwartzchild radius of the object. For such a source, we find a maximum energy of
\begin{equation}
E_{\rm{max}} \propto \gamma B M.
\end{equation}
With only micro-gauss galactic magnetic fields, we must turn to extragalactic sources to accelerate cosmic rays to energies above the EeV-scale. Extragalactic sources potentially capable of accelerating particles to such energies include the relativistic jets of Active Galactic Nuclei (AGN) or Gamma-Ray Bursts (GRB).

As protons are accelerated to very high energies in such sources, they may undergo photo-meson interactions with the surrounding radiation fields. In such interactions, both charged and neutral pions are produced. These pions then decay producing neutrinos and gamma-rays, respectively. This process essentially guarantees the existence of accompanying neutrinos and gamma-rays given the observed cosmic ray flux.
Alternatively, bounds can be placed on the cosmic neutrino flux by relating it to the cosmic 
ray spectrum.  
Using this method, Waxman and Bahcall have placed a upper bound 
for each neutrino flavor  \cite{wbbound}, 
\begin{equation}
E^2_{\nu}dN_{\nu}/dE_{\nu} \lsim 10^{-8}\  \rm{GeV}/(\rm{cm}^2\,\rm{s}\,\rm{sr}).
\label{wbbound}
\end{equation}
This assumes that the sources in question are optically thin, or transparent, to protons. If sources are optically thick to protons, howbever, the bound on the neutrino flux can only be based on gamma-ray observations by EGRET and is thus 
weaker by a factor of about 40 \cite{mpr}. Furthermore, if some sources were truly ``hidden'', meaning neither nucleons nor photons could escape, no upper bound could be made on the corresponding neutrino flux. For a further discussion of these and similar arguments, see Ref.~\cite{Albuquerque:2001jh}.

In addition to the neutrino production from cosmic ray interactions in or near cosmic accelerators, 
ultrahigh-energy cosmic rays produce neutrinos during propagation over cosmological distances \cite{cosmogenic}. Protons of energy above a few times $10^{19}$ eV can scatter off of Cosmic Microwave Background (CMB) photons with a center-of-mass energy roughly equal to the resonance corresponding to the mass of the $\Delta$-hadron (1.232 GeV). Again, both charged and neutral pions can be produced this way, yielding neutrinos and gamma-rays. The neutrino flux corresponding to this process is called the ``cosmogenic neutrino flux''. Unlike the flux of neutrinos from cosmic accelerators, the spectrum of cosmogenic neutrinos depends only on the spectrum of ultrahigh-energy protons and the distribution of their sources, and thus can be reliably calculated \cite{cosmogenicerrors}. The cosmogenic neutrino flux is often thought of as a ``guaranteed'' source of ultrahigh-energy neutrinos, assuming that the cosmic ray primaries at the highest observed energies are protons and not heavy nuclei \cite{Hooper:2004jc}.

There are ways in which neutrino fluxes larger than those described here could be produced. For example, in models of non-accelerator cosmic ray origins, {\it i.e.}, models in which the 
highest-energy cosmic rays are produced in the decay or annihilations of superheavy objects. In such models, the resulting neutrino flux is not constrained by the arguments shown here. Although such scenarios are certainly interesting, we will not consider them in this article. For further discussion on cosmic neutrino fluxes and their constraints, see Ref.~\cite{constraints}.

\begin{figure}[t]
\centering\leavevmode
\includegraphics[width=3.5in]{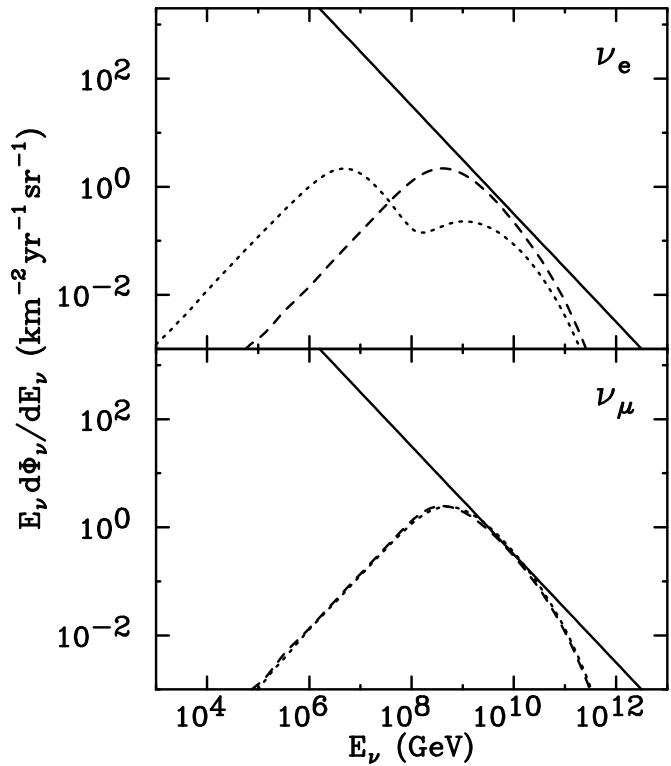}
\caption{The cosmic neutrino flux corresponding to the bound of Waxman and Bahcall (solid) \cite{wbbound} and the cosmogenic neutrino flux (dashed for $\nu$ and dotted for $\bar{\nu}$) \cite{cosmogeniccalc}. Note that after neutrino oscillations are considered, these fluxes will contain all three neutrino flavors ($\nu_e, \nu_{\mu}, \nu_{\tau}$) in equal quantities. The figure was taken from Ref.~\cite{bhicecube2}.}
\label{fluxes}
\end{figure}

Throughout the remainder of this article, we will primarily consider two representative choices for the cosmic neutrino spectrum. The first is a flux equal to the bound set by Waxman and Bahcall, which we call the Waxman-Bahcall flux of Eq.~(\ref{wbbound}). 
The second is the cosmogenic neutrino flux, as calculated in Ref.~\cite{cosmogeniccalc}. These are each shown in Figure~\ref{fluxes}.

\section{Particle Physics with High-Energy Neutrino Astronomy}
\label{part}

To identify potential signatures of new physics in high-energy neutrino interactions, one must first understand the phenomenology predicted by the Standard Model. In particular, the features of charged and neutral current interactions between high-energy neutrinos and target nuclei. The Standard Model predicts the cross sections for neutrino-nucleon interactions, up to uncertainties in parton distribution functions at extremely small values of the momentum fraction $x$, 
to energies beyond those probed by any planned neutrino telescopes \cite{neutrinocross}. In this section, we will describe the experimental features predicted by the Standard Model in neutrino telescopes and air shower experiments.  

\subsection{Neutrino Telescopes}

Neutrino telescopes are essentially arrays of detectors distributed over a large volume of a Cerenkov medium, such as water or ice. These detectors may be sensitive to optical Cerenkov radiation, as are AMANDA-II, IceCube and ANTARES, or radio, as is RICE. We will focus on optical Cerenkov detectors here, although the treatment of shower detection is quite easily generalized to include radio.

Muons produced in the charged current interactions of muon neutrinos can travel several kilometers through a detector medium producing a ``track" of Cerenkov light which can be observed and accurately reconstructed by neutrino telescopes.

The rate of muon events observed in a large volume neutrino telescope is given by 
\begin{equation}
\label{muonsrate}
N= \int  \int \int   dE_{\nu} \, dy \, d \cos \, \theta_z \, N_A \, 
\frac{dN_{\nu_{\mu}}(E_{\nu})}{dE_{\nu_{\mu}} dt \, d\Omega}\, 
R_{\mu}(E_{\mu}, \theta_z) \, \frac{d\sigma}{dy}(E_{\nu_{\mu}}) \, P_S(E_{\nu}, \theta_z) \,  A_{\rm{eff}} T,
\label{Nint}
\end{equation}
where $\theta_z$ is the zenith angle of an event ($\theta=0$ is vertically downgoing), $N_A$ is Avogadro's number, $dN_{\nu}/dE_{\nu} dt d\Omega$ is the flux of muon neutrinos 
(per unit energy, per unit time, per solid angle), $d\sigma/dy$ is the differential neutrino-nucleon cross section (where $y$ is defined such that $E_{\mu}=(1-y)E_{\nu_{\mu}}$), $P_S(E_{\nu}, \theta_z)$ is the survival probability of a neutrino travelling through the Earth, $A_{\rm{eff}}$ is the effective area of the detector ($\simeq 1$ km$^2$ for IceCube), $T$ is the length of time observed and $R_{\mu}(E_{\mu}, \theta_z)$ is either the muon range or the length of material ({\it i.e.} ice) between the detector and the Earth's surface, whichever is smaller. The muon range is defined as the distance a muon propagates in the medium surrounding the detector before falling below a cutoff energy. The muon range is given by \cite{range}
\begin{equation}
R_{\mu}= \frac{1}{\beta} \ln \bigg(\frac{\alpha+\beta E_{\mu}}{\alpha + \beta E_{\mu}^{\rm{cut}}}\bigg),
\end{equation}
where $E_{\mu}^{\rm{cut}}$ is the minimum muon energy required to produce an event. This value is selected to reduce the number of background events while retaining as many signal events as possible. In Optical Cerenkov neutrino telescopes, muons with energy as low as $10-100$ GeV can be observed, although cuts well above this energy are often imposed when searching for high-energy neutrinos. In ice, $\alpha \simeq 2 \times 10^{-6} \, \rm{TeV}\,\rm{cm}^2/\rm{g}$ and $\beta \simeq  4.2 \times 10^{-6} \,\rm{cm}^2/\rm{g}$ \cite{range}. For a PeV muon in ice, and a 100 GeV muon energy threshold, the range is approximately 1.7 km. For muons with energies of 10 PeV, 100 PeV or 1 EeV, the range increases to 7, 13  and 18 km, respectively. Thus for very energetic muon neutrinos, the target volume of the experiment becomes a long cylinder, rather than a box. This is particularly relevant for neutrinos coming from a direction near the horizon.

Unlike in accelerator experiments, the flux (or luminosity) of a cosmic neutrino beam may be unknown. Therefore, simply counting the number of events will not provide sufficient information to measure a neutrino cross section. Instead, information from the angular and energy distributions of events must be used \cite{crossmeasure}. For a cross section of  about $2 \times 10^{-7}\,$mb, 
a particle's interaction length as it travels through the Earth is equal to the Earth's diameter. This cross section is reached near $E_{\nu}\sim 100 \,$TeV according to the Standard Model prediction. Thus, as the neutrino-nucleon cross section is increased from its Standard Model value, the effect of absorption in the Earth becomes more pronounced and fewer of the observed events will come from neutrinos travelling through the Earth. A crude way to measure the neutrino-nucleon cross section could, therefore, be a comparison of the upgoing to downgoing (or Earth-skimming, etc.) events in a high-energy neutrino telescope. A more sophisticated analysis of the angular distribution of events as a function of energy would be more useful, however.

In addition to muon tracks, neutrino telescopes are sensitive to electromagnetic and hadronic showers. These events can be produced by all three neutrino flavors in neutral current interactions, or in some charged current interactions. For example,  electromagnetic showers are produced in the charged current interaction of an electron neutrino. The rate of shower events is calculated in an expression similar to Eq.~(\ref{muonsrate}), but the muon range, $R_{\mu}$, together with the effective area, $A_{\rm{eff}}$, are replaced by the effective volume of the detector. Also, the shower's energy is given by $E_{\rm{sh}}=y E_{\nu}$ for neutral current events and $E_{\rm{sh}}=E_{\nu}$ for electron neutrino charged current events. The minimum energy a shower must have to be observed by an optical Cerenkov neutrino telescope is on the order of a few TeV. For radio detectors, the shower threshold is much higher, in the PeV to EeV range.

Finally, very large volume neutrino telescopes, such as IceCube, are also capable of observing events uniquely associated with tau neutrinos. Tau neutrinos which interact via charged current in the detector medium produce a shower and a charged tau lepton. Below a few PeV, the tau lepton's lifetime is sufficiently short that it decays producing a second shower essentially spatially coincident with the first one. Such an event is indistinguishable from a single shower. At higher energies, however, the tau lifetime can be long enough to distinguish these two showers. For example, at 10 PeV, a tau travels, on average, about 500 meters before decaying.  If both showers occur within the detector volume, such an event is called a ``double bang'', and is a clear signature of a tau neutrino \cite{doublebang}. If the first of these showers occurs outside of the detector, with only the second shower being observed, the event is called a ``lollipop''. Here the observation of the shower with a minimum ionizing track (produced by the tau) constitutes the candy and the stick of the ``lollipop'', respectively. Again, this is a clear signature of a tau neutrino. 

Standard cosmic accelerators produce neutrinos via charged pion decay (see section~\ref{flux}). Pion decays produce flavors of neutrinos in the ratio $\phi_{\nu_{e}}:\phi_{\nu_{\mu}}:\phi_{\nu_{\tau}}=1:2:0$. Over the long baselines such neutrinos travel before reaching Earth, neutrino oscillations modify this ratio to $\phi_{\nu_{e}}:\phi_{\nu_{\mu}}:\phi_{\nu_{\tau}}\cong 1:1:1$, or nearly equal quantities of all three flavors. Considering only Standard Model neutrino interactions, these incoming flavor ratios can be translated to ratios of observed muon tracks, electromagnetic and hadronic showers and tau unique events \cite{flavormeasure}. By measuring the ratios of these event types observed in IceCube, the presence of interactions beyond the Standard Model may be tested.

\subsection{Neutrinos in Air Shower Experiments}

Very high-energy cosmic neutrinos can occasionally interact with particles in the Earth's atmosphere producing extended air showers observable in high-energy cosmic ray experiments such as AGASA, HiRes or the next generation Pierre Auger 
Observatory \cite{auger,reviewair}. Although the characteristics of neutrino induced showers do, in principle, differ from those initiated by hadronic cosmic rays \cite{profile}, significantly more hadronic showers are expected, thus making showers from neutrino primaries difficult to conclusively identify.

Primary particles which have a near-horizontal trajectory can provide an opportunity to distinguish neutrinos from hadronic events, however. In contrast to hadronic cosmic rays (which interact in the top of the atmosphere), neutrino primaries have considerably smaller cross sections and thus interact with nearly equal probability throughout the atmosphere. If a shower is observed which was initiated deep inside of the atmosphere, it can be associated with a neutrino (or other weakly interacting particle) primary. For there to be sufficient column depth to make this distinction (typically 3000 to 4000 g/cm$^2$ is required), only primaries within about $15^{\circ}$ of the horizon can be considered \cite{bhairother,fengair1}. The class of cosmic ray events which can be associated with neutrino primaries are called ``deeply penetrating, quasi-horizontal showers''.

To calculate the rate of neutrino induced deeply penetrating, quasi-horizontal showers in an air shower experiment, one must estimate the acceptance to neutrino detection. This quantity is essentially the effective target mass multiplied by the accessible solid angle. It is often given in units of km$^3$ water equivalent steradians (km$^3$ we sr), where, for example, 1 km$^3$ water equivalent would be the target mass contained in one cubic kilometer of water or ice. Ref.~\cite{fengair1} estimates the acceptance of the AGASA experiment to deeply penetrating, quasi-horizontal showers to be 0.05 km$^3$ we sr at $10^8$ GeV, rising to 1.0 km$^3$ we sr at $10^{10}$ GeV and above.  They estimate the acceptance of Auger to be a factor of 20, 20 and 50 larger than AGASA at $10^8$, $10^{10}$ and $10^{12}$ GeV, respectively. These estimates consider showers within 15$^\circ$ of the horizon and with a maximum height of 15 kilometers. For a discussion of the HiRes acceptance, see Ref.~\cite{hiresacceptance}.

The number of neutrino events observed as deeply penetrating, 
quasi-horizontal showers is given by
\begin{equation}
\label{bhrate}
N= \int dE_{\nu} N_A \frac{dN_{\nu}(E_{\nu})}{dE_{\nu} dt d\Omega} 
\sigma(E_{\nu}) A(E_{\nu}) T,
\label{Nair}
\end{equation}
where $N_A$ is Avogadro's number, $dN_{\nu}/dE_{\nu} dt d\Omega$ is the 
flux of neutrinos (per unit energy, per unit time, per solid angle),
$\sigma(E_\nu)$ is the neutrino-nucleon scattering cross section,
$A(E_{\nu})$ is the acceptance of the detector and $T$ is the length of time observed. This expression assumes that roughly all of the neutrino's energy goes into the produced shower. If this is not the case, such as with neutral current interactions, a differential cross section should be used as in Eq.~(\ref{muonsrate}) and the acceptance be written as a function of the shower energy rather than neutrino energy.

At the very high energies at which air shower experiments are most effective (0.1 EeV and higher), a reasonable and conservative flux of neutrinos to consider is the cosmogenic flux (see section~\ref{flux}). This flux peaks at about 0.1 EeV, but is substantial at 1 EeV and above. If we insert this flux, an experimental acceptance and the neutrino-nucleon cross section, we can predict the number of neutrino induced deeply penetrating, quasi-horizontal events which would be observed in an air shower experiment. 

In addition to deeply penetrating quasi-horizontal showers, it may be possible to identify showers produced by Earth-skimming tau neutrinos using the fluorescence detectors of the Auger experiment \cite{tauauger}. Earth-skimming, ultrahigh-energy tau neutrinos produce tau leptons in charged current interactions. Since at ultrahigh-energies the tau decay length is comparable to its interaction length, a shower produced in the tau decay can be observed as an extended air shower if the tau is produced not too deep beneath the Earth's surface. The rates for this class of events is expected to be rather small, however, and we will not study this signature in further detail here.

\section{Signatures of New Physics}
\label{theory}

In high-energy cosmic neutrino experiments, the new physics
enters the neutrino-nucleon scattering cross section, 
$\sigma_{\nu N}(E_\nu)$, 
as in Eqs.~(\ref{Nint}) and (\ref{Nair}). It is given by
\begin{equation}
\sigma_{\nu N}(E_\nu) = \sum_i \int_{\xmin}^1 dx\,
\hat{\sigma}_i ( xs ) \, f_i (x, Q) \ ,
\label{cross}
\end{equation}
where $\hat{\sigma}_i$ is the scattering cross section for the
neutrino-parton subprocesses, which reflects the fundamental dynamics
of neutrino interactions.
The sum is over all contributing partons, $i$, and $f_i$ are the parton 
distribution functions. The cross sections are insensitive to choice of
momentum transfer, $Q$. The cross sections are also insensitive to
uncertainties in the parton distribution functions at low $x$ if there is a TeV scale threshold.  For instance,
the highest-energy neutrino fluxes, 
which are large at $E_{\nu} \sim 10^{10}~\gev$, 
probe $x \sim (1~\tev)^2 / 10^{10}~\gev^2 \approx 10^{-4}$, 
within the range of validity of the parton
distribution functions that we take as CTEQ5~\cite{Lai:2000wy}.

There are many interesting scenarios in which neutrino-nucleon interactions 
would be substantially enhanced over the Standard Model prediction
at high energies. 
In this section, we summarize the literature on these models, 
describing how their signatures may  be 
observed in high-energy cosmic neutrino experiments.

\subsection{Contributions of KK Gravitons}
\label{kk}

The most dramatic proposal, perhaps, is that of TeV scale gravity with the existence 
of extra spatial dimensions. The motivation for such a scenario can be traced to the understanding
of the large hierarchy between the electroweak scale, below about $\cal O$(1 TeV), and the
Planck scale, $M_{PL}\sim 10^{19}$ GeV. If there are $n$ extra spatial dimensions,
then the 4-dimensional Planck scale is related to the $D$-dimensional Planck scale, $M_D$, by
\begin{equation}
 M_{PL}^2 \sim V_n M_D^{n+2},
\end{equation} 
where $V_n\sim R^n$ is the volume of the extra dimensions if they are flat with
a compactification scale, $R$ \cite{add}; and  $V_n\sim \int e^{-kx_i} d^nx$ 
if the $i^{th}$ extra dimension is ``warped" with a curvature $k$ \cite{rs}.
An immediate implication of
this scenario would be to naturally understand 
the largeness of the Planck scale in comparison with the electroweak scale.
Namely it can be interpreted as due to the
large volume of the extra dimensions while the fundamental 
 $D$-dimensional Planck scale, $M_D$, may be low, possibly at the TeV scale.

\begin{figure}[tb]
\centering\leavevmode
\includegraphics[width=4.0in]{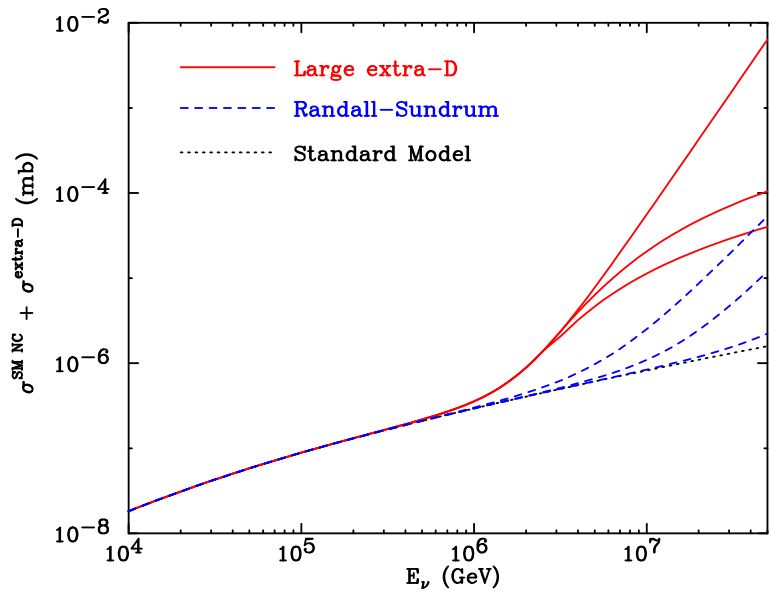}
\caption{The neutrino-nucleon cross section for models of TeV-scale gravity. 
The solid lines represent models with large extra dimensions (ADD scenario), with varying choices for the unitarization scheme of  the partial wave amplitudes  \cite{hangravity}. 
Short dashes lines represent Randall-Sundrum models with varied scales ($1-3$ TeV) and KK graviton masses (500 GeV$-1$ TeV).
Dotted lines represent the prediction of the Standard Model.}
\label{gravitysigma}
\end{figure}

In models with large and flat extra dimensions, often called the ADD scenario \cite{add}, 
the fundamental Planck scale is assumed to be 
on the order of a TeV and a large number of Kaluza-Klein (KK) 
graviton states of mass $1/R$ of sub-eV become accessible.
Their couplings to the Standard Model particles at an energy $E$ are  enhanced to
 $(E R)^n / M_{PL}^2 \sim E^n/ M_D^{n+2}$, and thus
potentially large effects on high-energy processes may occur \cite{addpheno}.
High-energy neutrinos can exchange these Kaluza-Klein gravitons with quarks or gluons in target nucleons resulting in an enhancement.
For a scale of gravity near 1 TeV, neutrinos above PeV energies begin to interact largely 
by the effects of new physics. 
In the Randall-Sundrum scenario, an anti-de Sitter dimension with a non-factorizable warped geometry is introduced \cite{rs}. Again, Kaluza-Klein gravitons become accessible at the TeV 
scale \cite{rspheno}
enhancing the neutral current neutrino interaction rate above this scale.  

\begin{figure}[tb] 
\centering\leavevmode
\includegraphics[width=2.7in,angle=90]{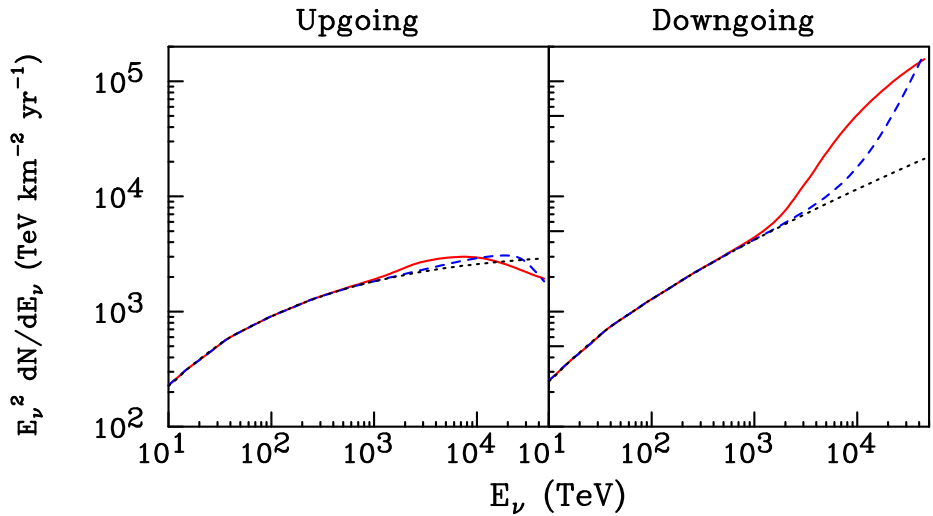}
\caption{The energy distribution of $\nu_\mu+\bar\nu_\mu$ events in Icecube. 
The left and right panels show the upgoing and downgoing events, respectively. The Waxman-Bahcall flux has been used (see section~\ref{flux}). In each panel, the solid line is the event rate  
when the neutrino-nucleon cross section 
$\sigma^{\rm SM}~+~\sigma^{\rm ADD}$ is used; the dashed line uses  
$\sigma^{\rm SM}~+~\sigma^{\rm RS}$  with a KK graviton mass $500$ GeV and  
$\Lambda=1$ TeV;  the dotted line corresponds to $\sigma^{\rm SM}$ alone.
The figure is a modified version from Ref.~\cite{hangravity}.}  
\label{two}
\end{figure}

The effect on the neutral current neutrino-nucleon cross section in these scenarios is 
shown in Fig.~\ref{gravitysigma}  \cite{hangravity}.  
The solid lines represent models with large extra dimensions (ADD)
for the quantum gravity scale taken as 1 TeV. The three curves correspond to different
choices for the unitarization scheme of the partial wave amplitudes proposed
in  \cite{hangravity}, as compared to some other calculations \cite{unitarity}.
Short dashed lines represent the Randall-Sundrum model with a varied 
AdS scale ($\Lambda=1-3$ TeV) 
and KK graviton masses (500 GeV$-1$ TeV).
Dotted lines represent the prediction of the Standard Model.
In the scenarios considered here, 
neutrino telescopes expect to observe more neutral current events per charged current event than predicted by the Standard Model. 
This effect sets in right above the threshold near $M_D$, providing
a clear indication for new physics.
Furthermore, the ratio of downgoing to upgoing events will be enhanced over the Standard Model prediction as more neutrinos are absorbed as they propagate though the Earth \cite{hangravity}.
The behavior of the energy spectrum due to these effects is shown in Fig.~\ref{two}. 
 
\subsection{TeV String Resonances}
\label{sr}

At energies above the compactication scale, $1/R$, the extra dimensions
and KK effects may become observable if $\md$ is not too high as discussed
in the previous section.  At even higher
energies near the string scale, $\ms$, string effects dominate over gravitational
effects based on string perturbation arguments \cite{res,shiu}. 
The string scale may be related to the $D$-dimensional gravity scale by
\begin{equation}
\ms = \kappa \md,
\end{equation}
with $\kappa\lsim 1$, depending on the string coupling and compactification of
the extra dimensions. Large effects from string resonances may be
produced when $E\gsim \ms$. 

\begin{figure}[tb]
\centering\leavevmode
\includegraphics[width=4.0in]{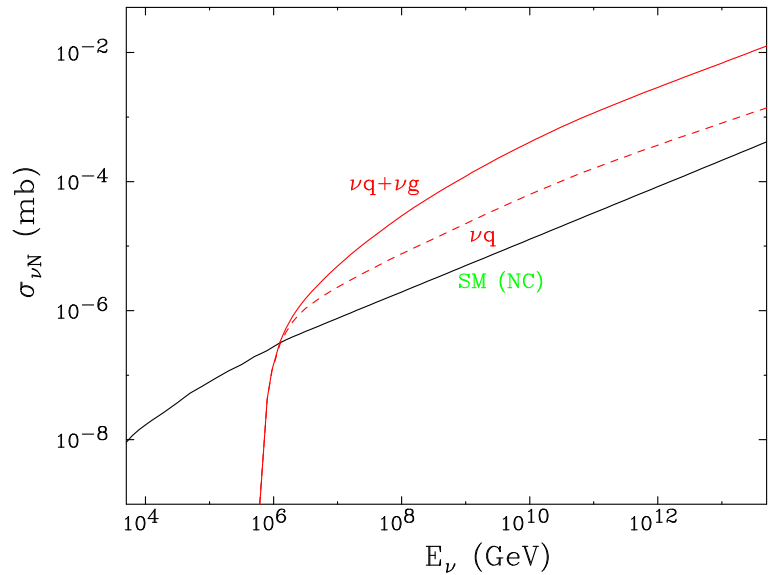}
\caption{$\nu N$ cross sections via TeV string resonances considering $\nu q$ and $\nu g$ contributions (dashed curved) \cite{stringy2}
and the $\nu q$ contribution only (dot-dashed curve).
The string scale is taken to be $\ms$=1 TeV and the 
Chan-Paton factor used is $T=1/2$.
Also plotted is the Standard Model neutral current prediction (solid curve). }
\label{stringsigma}
\end{figure}

General tree-level open-string amplitudes can be constructed based on 
the Veneziano amplitude \cite{res},
\begin{equation}
S(s,t) = \frac{\Gamma(1 - \alpha's)\Gamma(1 - \alpha't)}
{\Gamma(1 - \alpha's - \alpha't)} ,
\end{equation}
where $\alpha'=M_S^{-2}$ is the string tension. This amplitude develops simple
poles at $\sqrt s= \sqrt n \ms$ with $n=1,2, ...$ leading to resonances in the
matrix elements. 
The physical effects of these resonances have been explored  \cite{shiu,vene},
including their signatures in cosmic neutrino experiments \cite{stringy1,stringy2}.
We present the neutrino-nucleon cross sections due to Veneziano amplitude resonances in Fig.~\ref{stringsigma} \cite{stringy2}. The solid curve
shows the prediction for the Standard Model neutral current process, while
the dashed and dot-dashed curves represent string excitations with and without 
the gluon contribution, respectively.
We see that neutrino-gluon scattering can be  the dominant
process, 5 to 10 times larger than the neutrino-quark
induced processes. It is interesting to note that even for processes  
that vanish in the Standard Model at tree-level, there can still be
substantial stringy contributions to their amplitudes  at high energies.
Generally speaking, the energy-dependence of the cross sections
for the string resonances are weaker than those for KK states.

The event rates expected in IceCube and Auger have been calculated for these models \cite{stringy1,stringy2}. 
With the Standard Model interactions, 
only about 0.2 (0.7) shower events per year are expected in the  experiment from a cosmogenic 
(WB) neutrino flux,
These rates can be enhanced by a factor of 5 to 6 due to string excitations with 
$M_S=1$ TeV or a factor of about 1.5 to 1.7 with $M_S=2$ TeV. 


\subsection{Microscopic Black Hole Production}
\label{bh}

\begin{figure}[t]
\centering\leavevmode
\includegraphics[width=3.8in]{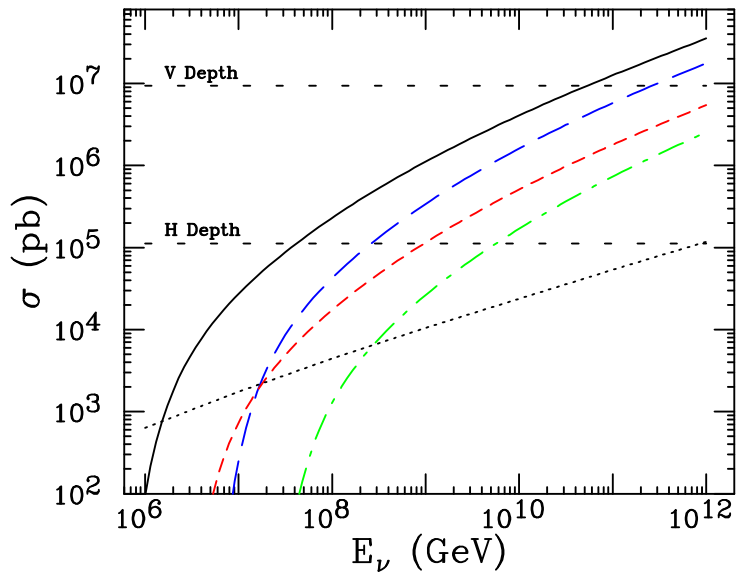}
\caption{The neutrino-nucleon cross section for black hole production in models of TeV-scale gravity. The lines represent models with a scale ($M_D$) of 1 TeV and a minimum black hole mass of 1 TeV (solid), a scale of 1 TeV and a minimum black hole mass of 3 TeV (long dashes), a scale of 2 TeV and a minimum black hole mass of 2 TeV (short dashes), and a scale of 2 TeV and a minimum black hole mass of 6 TeV (dot-dashes). The dotted line represents the prediction of the Standard Model. The number of extra dimensions was assumed to be six. The figure was taken from Ref.~\cite{bhicecube2}.}
\label{bhcrossfig}
\end{figure}

\begin{figure}[t]
\centering\leavevmode
\includegraphics[width=4.0in]{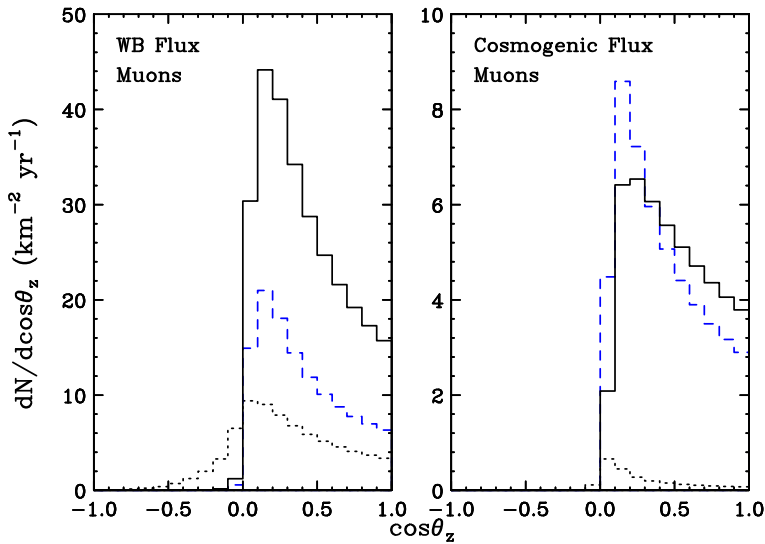}
\caption{The angular distribution of muon tracks above 500 TeV in a kilometer-scale neutrino telescope, such as IceCube, in models of black hole production~\cite{bhicecube2}. The dotted line represents the prediction for the Standard Model prediction while the solid and dashed lines are for the black hole production models with $x_{\rm{min}}=1$ and 3, respectively. All models shown have $n=1$ and $M_D=1\,$ TeV. The Waxman-Bahcall and cosmogenic neutrino fluxes were used in the left and right frames, respectively (see Section~\ref{flux}). $\cos \theta_z=0$ corresponds to a horizontal event while positive and negative values correspond to downgoing and upgoing muons, respectively. While the enhanced cross section dramatically increases the downgoing event rate, the rate of upgoing events is suppressed due to absorption in the Earth. The figure was taken from Ref.~\cite{bhicecube2}.}
\label{angle}
\end{figure}

At trans-Planckian energies, $E\gg M_D$, it has been argued that black hole production will be the leading process \cite{bhproduction,Kanti:2004nr}. This is because the energy-dependence of
the black hole production cross section grows faster than for sub-Planckian processes and
the number of non-perturbative states grows faster than the perturbative
string states. 
The cross section for black hole production can be naively estimated by the geometric description,
\begin{equation}
\label{parton}
\sigma(E_{\rm{CM}}) = \pi r_{\rm{sch}}^2(E_{\rm{CM}}),
\end{equation}
where $r_{\rm{sch}}(E_{\rm{CM}})$ is the Schwartzchild radius of a black hole formed with a mass equal to the center-of-mass energy of the collision. In $4+n$ dimensions, the Schwartzchild radius of a black hole of mass $M_{\rm{BH}}$ is given by
\begin{equation}
r_{\rm{sch}}(M_{\rm{BH}}) = \frac{1}{M_D} \bigg[\frac{M_{\rm{BH}}}{M_D} \bigg]^{1/(1+n)} \bigg[\frac{2^n \pi^{(n-3)/2}\,\Gamma(\frac{3+n}{2})}{2+n}\bigg]^{1/(1+n)}.
\end{equation}
Although some studies support the validity of the geometric cross section argument \cite{valid}, 
it is possible that a substantial fraction of the total energy will be radiated away in the form of gravitational waves, reducing the mass of any black hole which may be formed, and reducing the corresponding cross section \cite{fenginelastic}. 

Although the lowest mass possible for the black hole creation is approximately the fundamental Planck 
scale, $\md$, the effective center-of-mass energy should be several times larger for the
semi-classical argument to hold valid. 
 To parameterize this effect, we introduce the quantity, $x_{\rm{min}}= M^{\rm{min}}_{\rm{BH}}/M_D>1$. In addition to the ambiguity of the value of $x_{\rm{min}}$, other uncertainties can possibly arise in the estimation of the cross section \cite{cavaglia}, which we do not consider further here. 
 In Fig.~\ref{bhcrossfig}, 
 we show the cross section for black hole production in neutrino-nucleon interactions for different choices of $M_D$ and $x_{\rm{min}}$.
 As a result of the sum over all partons and the lack of suppression from small
perturbative couplings, the black hole cross section may exceed
Standard Model interaction rates by two or more orders of magnitude.
 The cross sections corresponding to neutrino interaction lengths equal to 
the horizontal and vertical depths of IceCube position 
are also given in Fig.~\ref{bhcrossfig} by the horizontal dotted lines.  We see that for the
geometric cross section, $\md \sim 1~\tev$, and neutrino energies
$E_{\nu} \sim 10^9~\gev$ where the cosmogenic flux peaks, 
black hole production increases the
probability of conversion in down-going neutrinos without increasing
the cross section so much that vertical neutrinos would be shadowed before
reaching  the detector.  
We therefore expect significantly enhanced rates in neutrino telescopes. 
The energy-dependence of the black hole production cross section is stronger than for the other processes we have discussed so far, thus confirming the argument that black hole production would likely be the dominant effect of low scale gravity at higher energies.

Black holes decay via Hawking evaporation almost instantly (with a  lifetime of the order 
$10^{-27}\, \rm{sec}$). The Hawking radiation follows a thermal distribution 
with temperature $T_H = (1+n)/4 \pi r_{\rm{sch}}$ with an average multiplicity of particles radiated of approximately $\langle N \rangle \cong M_{\rm{BH}}/2 T_H$. 
Naively, the particles are radiated in numbers proportional to their degrees of freedom, 
{\it i.e.}~$75\%$ hadronic, $10\%$ to charged leptons, etc.\footnote{The greybody
factors as the corrections for the black hole decays were calculated in Ref.~\cite{Kanti:2002nr}.
Also, specific models 
may give a somewhat different ratio of black hole decays when taking into account the constraint 
from nucleon stability \cite{Han:2002yy}.}
Assume that  the signals in a neutrino telescope
produced in black hole decays follow these ratios \cite{bhicecube1,bhicecube2}. The 10\% of Hawking radiation that produces charged leptons generates equal numbers of muons, taus and electrons which can be observed as muon tracks, tau unique events and electomagnetic showers, respectively. The full 75\% of Hawking radiation which goes into hadronic modes result in hadronic showers. This is in constrast to the ratios of event types predicted for Standard Model interactions. Taking into account the degrees of freedom corresponding to each channel and the factors effecting the probability of detection ({\it i.e.} muon range, etc.), the ratios of muons to taus to showers can be predicted for a particular black hole production model. For example, for a model with $M_D=1$ TeV and $x_{\rm{min}}=1$, about twice as many showers are expected than muon tracks (considering a $dN_{\nu}/dE_{\nu} \propto E_{\nu}^{-2}$ flux). In contrast, the Standard Model prediction is about 20\% more muons than showers \cite{bhicecube2}.
By combining flavor ratio measurements with angular and energy distributions, large volume neutrino telescopes such as IceCube will be capable of searching for evidence of black hole production in models with a fundamental Planck scale up to 1 to 2 TeV.

The angular distributions of muon tracks above 500 TeV in a kilometer-scale neutrino telescope, 
such as IceCube,  in models of black hole production are shown in Fig.~\ref{angle}.
While the enhanced cross section dramatically increases the downgoing event rate
($\cos \theta_z>0$) over the Standard Model prediction (dotted curves), 
the rate of upgoing events is suppressed due to absorption in the Earth.

Air shower experiments, unlike neutrino telescopes, do not have the ability to observe muon tracks or identify tau unique events. They are, however, very sensitive to EeV scale cosmic neutrinos and are thus capable of placing valuable limits on models of black hole production. Currently, the strongest such limit comes from the AGASA air shower experiment. AGASA has reported the observation of 1 neutrino-like (deeply penetrating, quasi-horizontal) event, and predict a background to this signal of 1.7 events from misidentifying hadronic primaries \cite{fengair1}. At the 95\% confidence level, this places an upper limit of 3.5 black hole events.  This can be directly translated into a limit on the fundamental Planck scale, $M_D$. For values of $x_{\rm{min}}$ in the range of $1-3$, AGASA can place a lower limit on $M_D$ of 1.0 to 1.4 TeV \cite{fengair1,fengair2}, a limit which is competitive to the strongest bounds from collider experiments \cite{bhcollider}. Auger, with considerably higher acceptance to these events, is expected to improve this sensitivity to $3-4$ TeV for $n \gsim 4$ \cite{fengair1}.

\subsection{$p$-Brane Production}
\label{pb}

\begin{figure}[tb]
\centering\leavevmode
\includegraphics[width=3in]{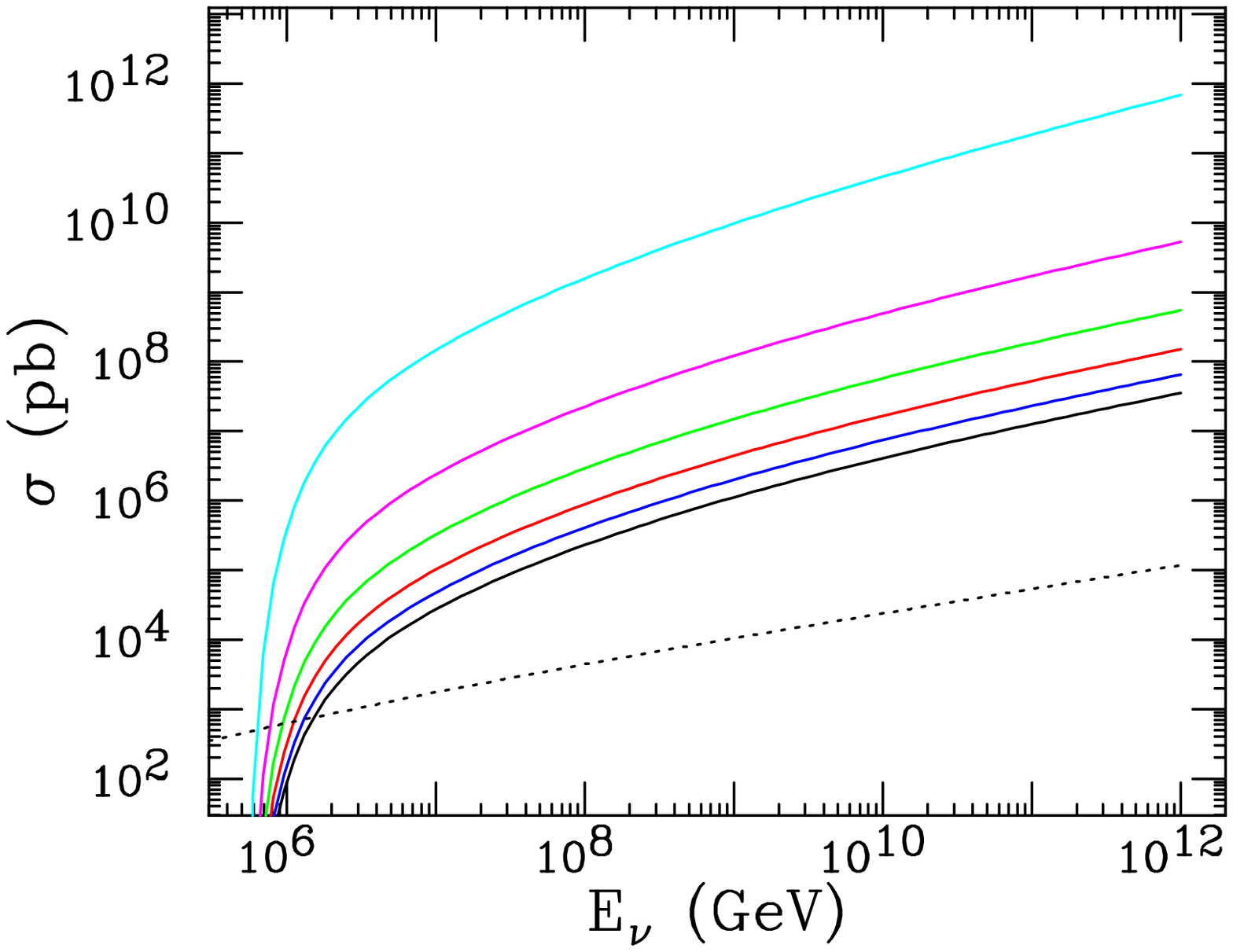}
\includegraphics[width=3in]{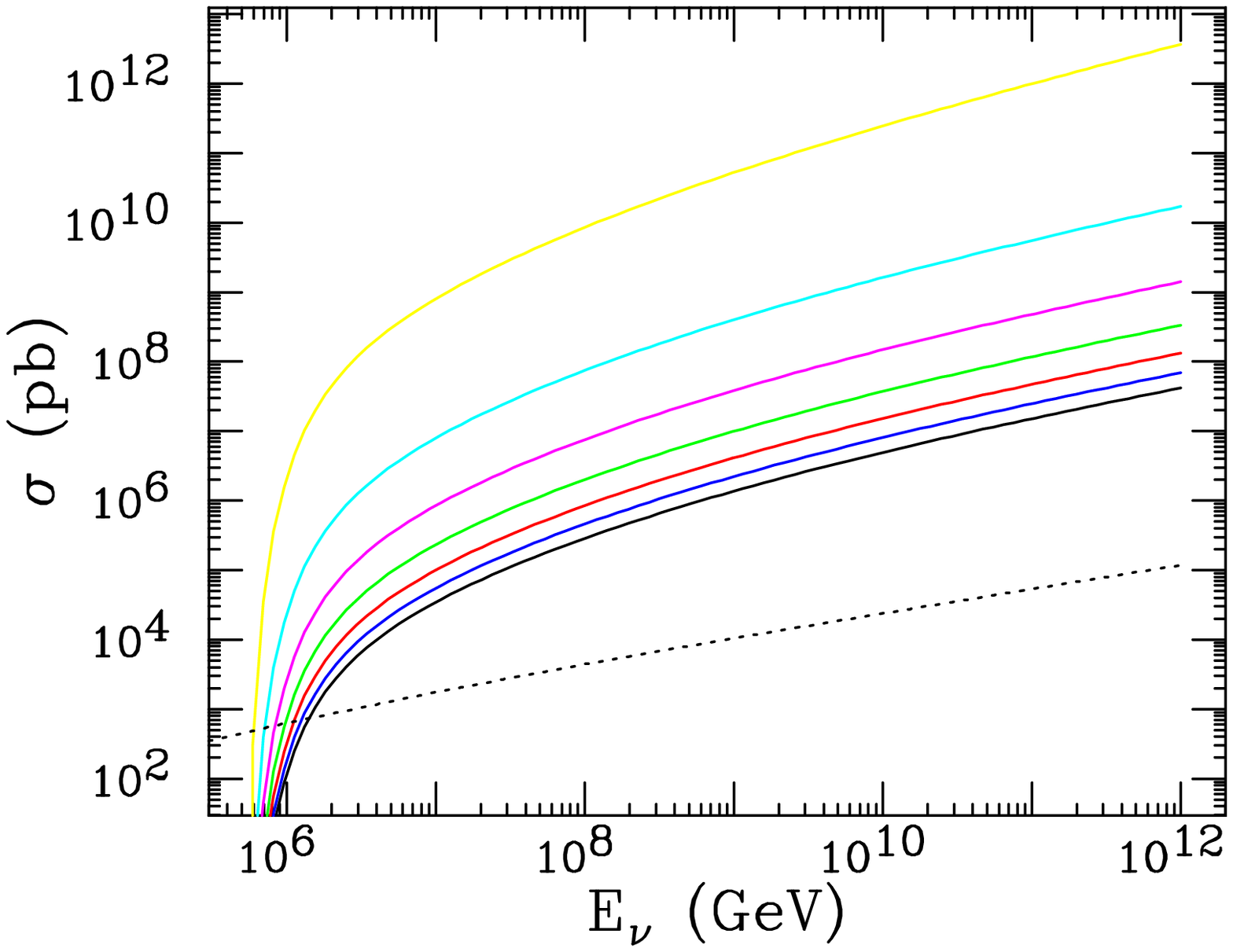}
\caption{The total cross section for $p$-brane production for 
$n=6$ (left), $n=7$ (right),  $\md= M_p^{min} =1~\tev$,
and $m = 0, ... , n-1$ from bottom to top. The Standard Model cross section,
$\sigma(\nu N \to \ell X)$, (dotted) is also shown. The figure was taken from 
Ref.~\cite{Anchordoqui:2002it}.}
\label{fig:sigma} 
\end{figure}

$p$-branes are $p$-dimensional, spatially extended solutions of gravitational
theories. The existence of such objects is a generic prediction of theories
with extra dimensions. If the fundamental scale of gravity is of the order of a TeV, then it is 
reasonable to expect that in addition to black holes (a  spherically symmetric $0$-brane), 
higher dimensional states may also be generated in high-energy 
collisions \cite{Ahn:2002mj,Cheung:2002uq}. 
The cross section for $p$-brane production is argued to be geometrical, similar to that for black hole production, except that it may have a lower threshold near the quantum gravity scale.
If the $p$-brane wraps only around the small (compact) dimensions, the cross section for $p$-brane production can be comparable to, or even larger than, the cross section for black hole production \cite{Cavaglia:2002si}.
If the p-brane wraps around large dimensions as well, their production will be suppressed by powers of 
$M_D/M_{\rm{PL}}$ \cite{Ahn:2002mj,Cheung:2002uq}. Typical cross sections for $p$-brane
production in $\nu N$ collisions are presented in Fig.~\ref{fig:sigma} \cite{Anchordoqui:2002it},
that could be higher than that of the black hole production by orders of magnitude. 

Unlike with the standard Hawking radiation picture for black hole evaporation, 
the decay of $p$-branes is not well understood.  $p$-branes may decay into branes of lower dimension. Alternatively, they may decay directly into a combination of brane and bulk particles. 
 
Below the energy threshold for $p$-brane or black hole production, lighter states, called ``string balls'' may also be produced~\cite{Dimopoulos:2001qe}. We do not study these objects further here, 
since we consider our presentation already quite representative for the conservative scenario
as in the string resonances in Sec.~\ref{sr}, and for the more optimistic scheme in this section.

\subsection{Electroweak Instanton Induced Processes}

Standard Model electroweak instantons represent tunnelling transitions between topologically inequivalent vacua, leading to baryon plus lepton number ($B+L$) violating processes. Such processes are exponentially suppressed below the so-called  ``Sphaleron'' energy, 
$E_{sph}\sim  \pi M_W/\alpha_W\sim 8$ TeV. 
Above this scale, however, such process may be 
unsuppressed and the corresponding cross sections can be quite large \cite{instanton},
potentially resulting in enhanced neutrino scattering signals in cosmic neutrino experiments.

\begin{figure}[t]
\centering\leavevmode
\includegraphics[width=3.0in,angle=90]{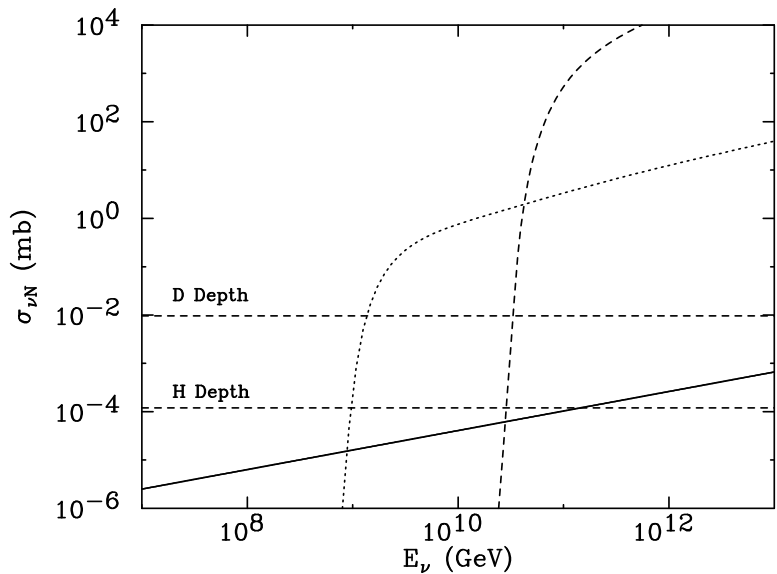}
\caption{The neutrino-nucleon cross section for electroweak instanton induced interactions. The dotted line represents the perturbative approach of Ref.~\cite{ringwald}. The dashed line represents the semi-classical approach of Ref.~\cite{rubikov}. The solid lines is the Standard Model neutral+charged current prediction. The figure was taken from Ref.~\cite{instantonhan}.}
\label{instantonsigma}
\end{figure}

The neutrino-nucleon cross section corresponding to electroweak instanton induced interactions is difficult to reliably calculate. One approach to this problem is to use a perturbative method in close analogy to QCD \cite{ringwald}. Alternatively, this calculation has been performed using a generalized semi-classical approach \cite{rubikov}. In this second approach, these interactions remain suppressed to much higher energies up to about 30 times the Sphaleron energy. 
The estimated neutrino-nucleon cross sections corresponding to these two approaches 
are shown in Fig.~\ref{instantonsigma} \cite{instantonhan}.

\begin{figure}[tb]
\centering\leavevmode
\includegraphics[width=3.0in,angle=90]{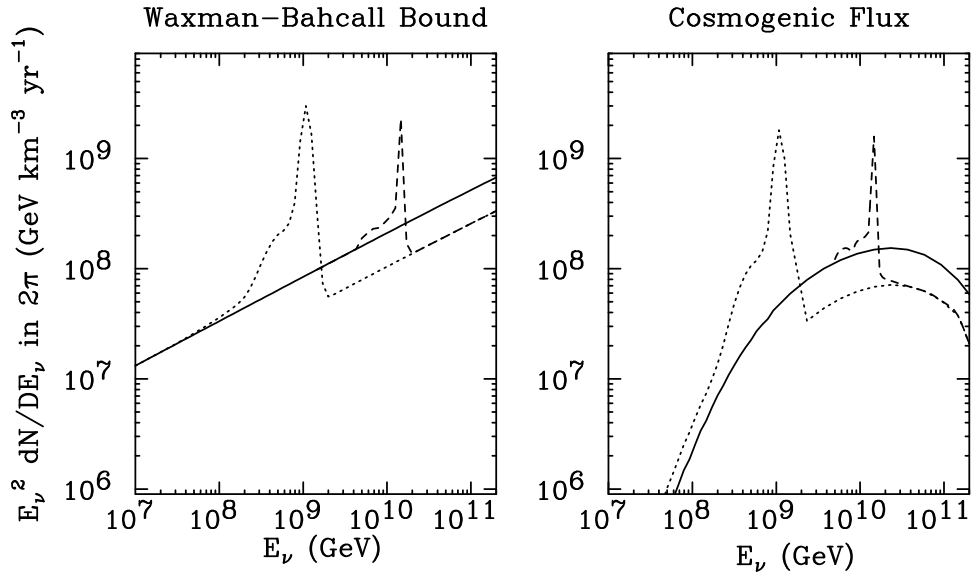}
\caption{The spectrum of neutrino (shower) events predicted in a neutrino telescope including the electroweak instanton induced interactions for the models described in the text. The dotted line represents the perturbative approach of Ref.~\cite{ringwald}. The dashed line represents the semi-classical approach of Ref.~\cite{rubikov}. The solid lines is the Standard Model prediction. The figure was taken from Ref.~\cite{instantonhan}.}
\label{instantonenergy}
\end{figure}

From the standpoint of neutrino phenomenology, it is important to note the extremely rapid increase of the neutrino-nucleon cross section demonstrated in these models. This is in contrast to the more gradual growth predicted for the cross sections for black hole production, Kaluza-Klein exchanges, etc. Below the energy thresholds for such interactions, the Standard Model predictions are accurate. At energies roughly a factor of ten higher, the cross section becomes sufficiently large that the Earth efficiently absorbs the incoming neutrino flux, as indicated above the horizontal dotted 
lines (horizontal and downward depths) in Fig.~\ref{instantonsigma}. 
Thus in a neutrino telescope, a sharp enhancement in a fairly narrow range of energies is predicted for these models, as depicted in Fig.~\ref{instantonenergy}. Although spectacular, 
the ability of planned experiments to observe such features is limited, however. Even with the more optimistic of the models considered here, an experiment such as IceCube is expected to see only on the order of one event per year from instanton induced processes~\cite{instantonhan}. 
Future  experiments with very large volume 
will be required to further probe such models.

Another interesting characteristic feature of instanton-induced processes
is the large multiplicity of final state particles and the violation of $B+L$.
The basic operators involving quark and lepton fields are of the 
form $\langle (qqq\ell)^{n_g} \rangle$ \cite{instanton}, where $n_g=3$ is
the number of fermion generations. It has been argued that the processes
involving multiple gauge bosons and Higgs bosons, 
such as $\langle (qqq\ell)^{n_g} W^n H^m \rangle$, can be significantly
enhanced \cite{highE}. A typical neutrino-induced event could thus be
\begin{equation}
\nu_e u \to \bar d \bar d\  
+ \ \bar c \bar c\bar s \mu^+
+ \ \bar t \bar t\bar b \tau^+\
+ nW + mH.
\end{equation}
With both quarks and leptons of all three generations involved simultaneously
in the primary production, this type of events should look quite unique. 
It is difficult to predict how such events would appear to the IceCube detector, however, given the fact that the particles will be highly collimated and difficult to separate.

\subsection{Signatures of Supersymmetry}

Supersymmetry (SUSY) remains a leading candidate for physics beyond the Standard Model. Although weak-scale SUSY is only weakly coupled to the Standard Model and generally would not lead to 
substantially enhanced neutrino scattering cross sections, certain charged particles 
produced by cosmic neutrinos  may be long-lived and may provide observable signatures. 
This scenario could be naturally realized when the gravitino is the stable 
Lightest Supersymmetric Particle (LSP), and a charged slepton (such as stau) is 
the Next-to-Lightest Supersymmetric Particle (NLSP)  \cite{susypair,supergravitygravitino}. 
Interactions of high-energy neutrinos may be able to produce pairs of sparticles 
which rapidly decay to charged slepton NLSPs which can only decay further into states including a gravitino. With only highly suppressed couplings allowing this decay, the NLSP stau can be sufficiently long lived to be potentially observable in a large volume neutrino telescope such as IceCube \cite{susypair}.

\begin{figure}[tb]
\centering\leavevmode
\includegraphics[width=3.0in,angle=90]{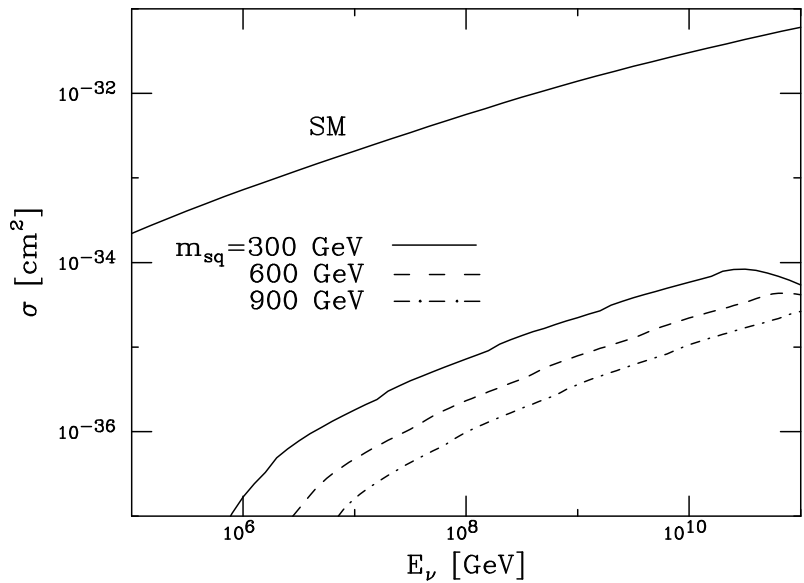}
\caption{The $\nu N$ cross section in the scenario of stau-NLSP. 
Curves are shown for squarks of mass 250, 600 and 900 GeV. A stau mass of 250 GeV was used. The chargino mass used was also 250 GeV.
The top curve corresponds to the Standard Model charged current interactions. The figure was taken from Ref.~\cite{susypair}.}
\label{fig:stau}
\end{figure}
 
Sparticle pair production in neutrino-nucleon interactions is dominated by 
the $t$-channel chargino exchange, resulting in a slepton and a squark. The squark then quickly decays into a slepton NLSP. The cross section for this process is rather small, however, typically two to three orders of magnitude below the Standard Model processes in the energy range well above the kinematic threshold as seen Fig.~\ref{fig:stau}.
The key observation here is that 
sleptons produced in neutrino interactions travel through the Earth, losing energy via ionization processes and radiation. Due to their much greater mass,  sleptons lose far less energy than muons produced in Standard Model charged current interactions. The ``slepton range'' can extend to hundreds or thousands of kilometers, thus in part making up for the low cross section for their production.

The two sparticles produced in these interactions travel from their point-of-origin separated by an angle of $\theta \simeq 2 m_{\tilde{l}}/E_{\nu}$. Therefore, the signature of this process consists of a pair of Cerenkov tracks, separated by a distance $L \theta$, where $L$ is the distance between the detector and the sparticles' point-of-origin. Considering a PeV neutrino, for example, two sleptons separated by $\theta \simeq  10^{-3}-10^{-4}$ could be produced. After travelling $\sim 1000$ kilometers, their tracks would be separated by $\sim 100-1000$ meters, which could be potentially distinguished in a neutrino telescope.  This ``double track'' signature would provide a method of distinguishing sparticle tracks from ordinary muon tracks.

Typical models with stau-NLPS predict only a one- or a few double-track events per year in 
IceCube \cite{susypair}.
Larger volume detectors, such as extensions of IceCube, may be needed to further explore this possibility.

As a final remark on 
the possibility of observing supersymmetric particles in high-energy cosmic-ray interactions, 
neutrino experiments may be able to identify  particles which are part of the cosmic ray spectrum, 
such as in cosmic ray models of top-down origin \cite{susysky}. For instance, 
if a neutralino is the LSP, it will interact with nucleons in a manner which somewhat 
resembles a neutrino neutral current interaction. Thus without very large fluxes of high-energy cosmic neutralinos, it would be very difficult to distinguish any such particles from neutrinos. 
Neutralinos can have considerably smaller cross sections with nucleons than neutrinos, however, allowing them to travel through the Earth at energies at which neutrinos will be efficiently absorbed \cite{susysky}.  Thus 
ultrahigh-energy, neutralino-induced showers provide a low background signal in the direction of the Earth. Future space-based air shower experiments, such as OWL or EUSO, may be sensitive to this signature in some scenarios \cite{susysky,Anchordoqui:2004qh}. 

\section{Conclusions and Summary}
\label{conclude}

In this article, we have reviewed the ability of neutrino telescopes and  air shower experiments to study particle physics with high-energy cosmic neutrinos. 
The main advantage of such experiments over traditional collider experiments is the higher
energy at which interactions can be studied. Several of the experiments we describe in this 
article could expect multiple events per year at energies above 1 EeV,  corresponding to 
about 40 TeV in the center-of-mass frame of a neutrino-nucleon collision. 
Clearly collider experiments have advantages over astroparticle techniques as well.  Most notably, the high luminosities and well-controlled conditions of collider experiments are luxuries astronomers do not often enjoy.

Together, these advantages and disadvantages determine the areas of particle physics in which neutrino astronomy can be most useful.  In particular, models in which dramatic deviations from the Standard Model occur at energies beyond the reach of colliders can often be tested in such experiments. In this article, we have reviewed several of such scenarios,  summarized as follows:

\begin{itemize}
\item  Kaluza-Klein gravitons in low-scale quantum gravity scenarios: 
sensitive near the threshold, presumably $E_{CM}\approx M_D \sim 1$  TeV or $E_\nu\sim 1$ PeV.
\item  TeV string resonances in low string-scale scenarios: 
near and above the string scale presumably $\ms\sim 1$ TeV, thus $E_\nu > 1$ PeV.

\item  $p$-brane production in low-scale quantum gravity scenarios: 
near and above the quantum gravity scale, presumably $\md\sim 1$ TeV, thus $E_\nu > 1$ PeV.
\item  Black hole production in low-scale quantum gravity scenarios: likely the dominant
signature at trans-Planckian energies  $E_{CM}\gg M_D \sim 1$ TeV, thus $E_\nu \gg 1$ PeV.

\item  Electroweak instanton induced processes: above the sphaleron energy
 $E_{CM} > 10$ TeV,  thus $E_\nu > 100$ PeV.

\item  Supersymmetry with a charged slepton as a long-lived NLSP. 
\end{itemize}

Although we did not discuss them in this article, other probes of exotic particle physics are possible using neutrino astronomy. These include searches for particle dark matter and neutrinos associated with models of top-down cosmic ray origin. 

Although astronomy, and not particle physics, is the primary objective of neutrino telescopes and air shower experiments, upcoming experiments such as IceCube, AUGER, ANITA, OWL and EUSO will each study interactions at energies well beyond the reach of colliders and provide complementary probes to the traditional techniques used to study our Universe at the smallest scales.

\section*{Acknowledgments}

We would like to thank Jaime Alvarez-Muniz for his assistance with the figures and Jonathan Feng for helpful comments. TH is supported in part by DOE grant No.~DE-FG02-95ER40896 and the Wisconsin Alumni Research Foundation. DH is supported by the Leverhulme trust.

\end{document}